\documentstyle[epsfig,ichep]{article}

\def\lapproxeq{\lower .7ex\hbox{$\;\stackrel{\textstyle <}{\sim}\;$}}
\def\gapproxeq{\lower .7ex\hbox{$\;\stackrel{\textstyle >}{\sim}\;$}}

\begin{document}

\title{MRS(1994): parton distributions of the proton} 
\author{A.D.\ Martin$^{\dag}$,R.G.\ Roberts$^{\S}$ and W.J.\ Stirling$^{\dag}$}

\address{${\dag}$Department of Physics, University of Durham, Durham DH1 3LE, 
England. \\ 
$^{\S}$Rutherford Appleton Laboratory, Chilton OX11 0QX, England.}   

\abstract{To obtain improved parton densities of the proton, we present a new 
global analysis of deep-inelastic and related hard scattering data including, in 
particular, the recent measurements of $F_2$ at HERA, of the asymmetry of the 
rapidity distributions of $W^{\pm}$ production at the FNAL $p\bar{p}$ collider 
and of the asymmetry in Drell-Yan production in $pp$ and $pn$ collisions.
We discuss the evolution of the new partons to low values of $Q^2$.} 

\twocolumn[\maketitle] 
The increase in the precision of deep-inelastic and related data over the last 
few years has led to a considerable improvement in our knowledge of the parton 
distributions of the proton, at least for $x \gapproxeq 10^{-2}$.  Here we 
present an updated analysis \cite{MRSA} which incorporates new recent data.   
The new data are (i) the H1 \cite{H1} and ZEUS \cite{ZEUS} measurements of 
$F_2$, which mainly constrain the sea quarks in the previously unexplored small 
$x$ domain ($x \lapproxeq 10^{-3}$), (ii) the measurement of the asymmetry of 
Drell-Yan production in $pp$ and $pn$ collisions by NA51 \cite{NA51} 
\begin{equation} 
A_{DY} = \frac{\sigma_{pp}-\sigma_{pn}}{\sigma_{pp}+\sigma_{pn}} = -0.09 \pm 
0.02 \pm 0.025 
\end{equation} 
at $x = \sqrt{\tau} = 0.18$ and, (iii) the asymmetry of the rapidity 
distributions of the charged leptons from $W^{\pm} \rightarrow \ell^{\pm}\nu$ 
decays by CDF \cite{CDF} 

\begin{equation} 
A(y_{\ell}) = 
\frac{\sigma(\ell^+)-\sigma(\ell^-)}{\sigma(\ell^+)+\sigma(\ell^-)} . 
\end{equation} 
These asymmetry data are shown in Figs.\ 1 and 2 respectively, together with the 
predictions of our previous global analysis \cite{MRSH} and the equivalent set 
of CTEQ partons \cite{CTEQ}.  Neither set gives a satisfactory description of 
both asymmetries.  The deficiency is not surprising.  The high precision muon 
and neutrino deep-inelastic structure function data (of BCDMS, NMC and CCFR), 
which provide the core constraints of the global analyses, determine essentially 
the parton combinations $u+d$, $\bar{u}+\bar{d}$ and $u+\bar{u}$, but do not pin 
down the remaining combination $\bar{d}-\bar{u}$.  Indeed the Drell-Yan 
asymmetry experiment was proposed \cite{ES} just because it was uniquely 
equipped to determine this combination.  The $W^{\pm}$ rapidity asymmetry is 
proportional to the slope of $u/d$ at $x = M_W/\sqrt{s} \simeq 0.05$ at 
Fermilab.  The asymmetry data therefore offer a fine-tuning of the $u,d,\bar{u}$ 
and $\bar{d}$ parton densities in the region $x \sim 0.1$.  

\begin{figure}[b]
\begin{center}
\end{center}
\caption{The Drell-Yan asymmetry in $pp$ and $pn$ 
collisions.  The MRS(H) and CTEQ2M curves pre-date the NA51 measurement, 
(1), whereas it is included in the MRS(A) global fit.}
\label{1}
\end{figure}
\vfill
\eject

\begin{figure}[t]
\begin{center}
\end{center}
\caption{The $W^{\pm} \rightarrow \ell^{\pm}\nu$ rapidity 
asymmetry, (2), measured by [5], together with the next-to-leading 
order descriptions obtained using MRS(H), CTEQ2M and MRS(A) partons.  
These data are included in the MRS(A) fit.}
\label{2}
\end{figure}

Our basic procedure is to parametrize the parton distributions $f_i$ at a 
sufficiently large $Q^2_0$ ($Q^2_0 = 4$ GeV$^2$) so that $f_i(x,Q^2)$ can be 
calculated at higher $Q^2$ using next-to-leading order Altarelli-Parisi 
evolution equations.  An excellent overall description of the data is obtained 
with the following simple parametrization  
\begin{eqnarray} 
xu_{\rm v} & = & A_u x^{\eta_1}(1-x)^{\eta_2} (1 + \epsilon_u \sqrt{x} + 
\gamma_u x) \nonumber \\ 
xd_{\rm v} & = & A_d x^{\eta_3}(1-x)^{\eta_4} (1 + \epsilon_d \sqrt{x} + 
\gamma_dx) \nonumber \\ 
xS & = & A_S x^{-\lambda}(1-x)^{\eta_S}(1 + \epsilon_S \sqrt{x} + \gamma_Sx) 
\nonumber \\ 
xg & = & A_g x^{-\lambda}(1-x)^{\eta_g} (1 + \gamma_gx) , 
\end{eqnarray} 
where the valence distributions $u_{\rm v} \equiv u-\bar{u}$ and $d_{\rm v} 
\equiv d-\bar{d}$, and where the total sea distribution $S \equiv 
2(\bar{u}+\bar{d}+\bar{s}+\bar{c})$.  We assume that $s = \bar{s}$.  At present 
there are not enough experimental constraints on the gluon to justify the 
introduction of an extra parameter $\epsilon_g$ in $xg$, or to determine the 
exponent $\lambda$ independent of that of the sea-quark distribution $S$. Three 
of the four $A_i$ coefficients are determined by the momentum and flavour sum 
rules.
The distributions are defined in the $\overline{{\rm MS}}$ renormalisation and 
factorization scheme and the QCD scale parameter $\Lambda_{\overline{{\rm 
MS}}}(n_f = 4)$ is taken as a free parameter.  

The flavour structure of the quark sea is taken to be 
\begin{eqnarray} 
2\bar{u} & = & 0.4(1 - \delta)S - \Delta \nonumber \\ 
2\bar{d} & = & 0.4(1 - \delta)S + \Delta \nonumber  \\ 
2\bar{s} & = & 0.2(1 - \delta)S \nonumber  \\ 
2\bar{c} & = & \delta S
\end{eqnarray}  
at $Q^2 = Q^2_0 = 4$ GeV$^2$, with 
\begin{equation} 
x\Delta \; \equiv \; x(\bar{d}-\bar{u}) \; = \; A_{\Delta} x^{0.4} 
(1-x)^{\eta_S} (1 + \gamma_{\Delta}x) . 
\end{equation}
The first hint that the $u,d$ flavour symmetry of the sea is broken (with 
$\bar{d} > \bar{u}$ on average) came from the evaluation of the Gottfried sum by 
NMC \cite{NMC}.  Now the NA51 Drell-Yan asymmetry measurement, (1), provides 
further evidence that $\bar{d} > \bar{u}$, which we allow through the 
parametrization of $\Delta$.  The 50\% suppression assumed for the strange sea 
in (4) is in excellent agreement with the CCFR next-to-leading order analysis 
\cite{CCFR} of their $\nu N \rightarrow \mu^-\mu^+X$ data.  The parameter 
$\delta$ in the charm sea in (4) is adjusted to reproduce the EMC deep-inelastic 
$F^c_2$ data \cite{EMC}, as explained in \cite{MRSA}.  

Apart from the CCFR neutrino structure function measurements at $x = 0.015$ and 
0.045, the new MRS(A) analysis gives a good description of all deep-inelastic 
and related data.  In particular it leads to a much improved description of the 
Drell-Yan asymmetry (Fig.\ 1) with relatively modest changes in the partons (see 
Fig.\ 3). The values of the parameters are listed in Table 2 of \cite{MRSA}. 
>From the highly constrained overall fit we conclude that the valence 
and sea quarks are well determined for $0.02 \lapproxeq x \lapproxeq 0.6$.  

\begin{figure}[h]
\begin{center}
\end{center}
\caption{A comparison at $Q^2 = 20$ GeV$^2$ of the new
MRS(A) partons [1] and the earlier MRS(H) partons [6].}
\label{3}
\end{figure}

The HERA measurements of $F_2$ \cite{H1,ZEUS} are the only constraint on the 
parameter $\lambda$ in (3) which controls the small $x$ behaviour of the sea $xS \sim 
x^{-\lambda}$.  In both the MRS(A) and MRS(H) analyses an excellent description 
of the HERA data is obtained with $\lambda = 0.3$, see, for example, Fig.\ 4.  
However the parameters $\lambda$ and $\epsilon_S$ in (3) are highly correlated 
and it is possible to obtain acceptable fits for $0.2 \lapproxeq \lambda 
\lapproxeq 0.4$. 

\begin{figure}[t]
\begin{center}
\end{center}
\caption{The description of the $F_2$ structure function measurements at $Q^2 = 
25$ GeV$^2$ by MRS(A) partons.  The updated HERA data [2,3] are shown.} 
\label{4}
\end{figure}

Since the sea quarks are driven by the gluon, via $g \rightarrow q\bar{q}$, we 
have assumed a common $x^{-\lambda}$ behaviour at small $x$.  There is, as yet, 
no experimental confirmation of this assumption, and indeed the ambiguity in the 
gluon distribution is by far the largest uncertainty in the parton densities.  
The only information on the gluon is (i) that it carries about 43\% of the 
proton's momentum at $Q^2 = 4$ GeV$^2$, (ii) that its value at $x \sim 0.3-0.4$ 
is constrained by the WA70 \cite{WA70} and UA6 \cite{UA6} measurements of large 
$p_T$ prompt photon production, $pp \rightarrow \gamma X$, and (iii) through its 
influence on the observed scaling violations in the structure function data, 
although here there is a correlation with the value found for the QCD coupling.  
We stress that all the recent parton analyses are global \lq\lq best fits" to 
the data and so the spread of the gluons obtained underestimates the true 
uncertainty in its distribution.  

\begin{figure}[h]
\begin{center}
\end{center}
\caption{The description of sample E665 [16], NMC [17] and updated H1 [2]
and ZEUS [3] $F_2$ data by MRS(A) partons 
modified as in (6).  The E665 data are preliminary and in the global low $Q^2$ 
fit [14] they have been renormalized upwards by 20\% to obtain consistency with 
the NMC measurements of $F_2$.} 
\label{5}
\end{figure}

Finally, we extrapolate the MRS(A) partons to low $Q^2$ and attempt to describe 
E665, NMC and SLAC data for $F_2$ down to $Q^2 = 0.5$ GeV$^2$ \cite{MRSQ}.  
As expected it is necessary to suppress the distributions by incorporating the 
theoretical requirement that they vanish as $Q^2$ as $Q^2 \rightarrow 0$.
We find a satisfactory description of the $F_2$ data can be obtained if the 
parton distributions are modified as follows 
\begin{equation} 
f_i(x,Q^2) \longrightarrow \frac{Q^2}{Q^2+M(x)^2} f_i(x,Q^2) 
\end{equation} 
where 
\begin{equation} 
M^2 = 0.015 \; {\rm exp}[1.54 \sqrt{\ell n(1/x)}] 
\end{equation} 
or, equally acceptable, 
\begin{equation} 
M^2 = 0.07 \; x^{-0.37}
\end{equation} 
in GeV$^2$. That is $M^2$ increases from about 0.15 GeV$^2$ at $x = 0.1$ to 
about 0.9 GeV$^2$ at $x = 0.001$.  In other words, the suppression sets in at 
higher values of $Q^2$, $Q^2 \sim M^2$, as $x$ decreases.  This trend is expected on 
theoretical grounds since, with decreasing $x$, the onset of shadowing 
corrections  is predicted to occur at higher values of $Q^2$ \cite{GLR}.  Fig.\ 
5 shows the description of a sample of the low $Q^2$ data, together with the
latest H1 and ZEUS measurements of $F_2$.

\end{document}